\documentclass[a4]{article}
\usepackage[left=2cm,right=2cm,top=2cm,bottom=2cm,bindingoffset=0cm]{geometry}
\usepackage{epsfig}
\usepackage{multicol}
\usepackage{epstopdf}
\usepackage{multirow}

\newcommand{\Ref}[1]{(\ref{#1})}
\newcommand{\beao}{\begin{eqnarray*}}
\newcommand{\eeao}{\end{eqnarray*}}
\newcommand{\be}{\begin{equation}}
\newcommand{\ee}{\end{equation}}
\newcommand{\bea}{\begin{eqnarray}}
\newcommand{\eea}{\end{eqnarray}}
\newcommand{\beq}{\begin{eqnarray}}
\newcommand{\eeq}{\end{eqnarray}}
\newcommand{\nn}{\nonumber}
\newcommand{\pa}{\partial}

\newcommand{\la}{\lambda}

\begin{document}
\title{ Magnetized  quark-gluon plasma at the LHC }
\author{
V.~ Skalozub\thanks{e-mail: Skalozubv@daad-alumni.de}\\
{\small Oles Honchar Dnipropetrovsk National University, 49010 Dnipro, Ukraine}\\
\\
P.~ Minaiev\thanks{e-mail: Minaevp9595@gmail.com}\\
{\small Oles Honchar Dnipropetrovsk National University, 49010
Dnipro, Ukraine}}

\date{}
\maketitle

\begin{abstract}
In QCD,   the strengths of the  large scale temperature dependent
chromomagnetic,  $B_3, B_8$, and usual magnetic, $H,$ fields
spontaneously generated in quark-gluon plasma after the
deconfinement phase transition ($DPT$), are estimated. The
consistent at high temperature effective potential accounting for
the one-loop plus daisy diagrams is used. The heavy ion collisions
at the LHC and temperatures $T $ not much higher than  the phase
transition temperature $T_d$ are considered.

 The critical temperature for the magnetized
plasma is found to be   $T_d(H) \sim 110 - 120$ MeV.  This is
essentially lower compared to  the zero field value $T_d(H=0)\sim
160-180$ MeV usually discussed in the literature. Due to
contribution of quarks, the color magnetic fields act as the
sources generating $H$. The strengths of the fields are $B_3(T),
B_8(T) \sim 10^{18} - 10^{19}G$, $H(T) \sim 10^{16} - 10^{17}G$
for temperatures $T \sim 160 - 220 $ MeV. At temperatures $T < 110
- 120$ MeV the
 effective potential minimum value being negative approaches to
zero. This is signaling  the absence of the background fields
 and color confinement.

\end{abstract}
\begin{multicols}{2}
\section{Introduction}
At the LHC experiments,  in heavy ion collisions a new matter
phase - quark-gluon plasma ($QGP$) - has to be produced. The
deconfinement phase transition temperature is expected to be of
order $T_d \sim $ 180 - 200 MeV. In theory, investigation of the
$DPT$ and $QGP$ properties were carried out by different method -
analytic perturbative and nonperturbative, various numerical
methods and Monte-Carlo simulations  on a lattice (see, for
example, \cite{Satz2012} - \cite{szab14-LATTICE2013-014}).
 $QGP$ and strong magnetic fields had been existed in the hot Universe
\cite{gras00-348-163}, \cite{eliz12-72-1968}.

 One of
distinguishable properties of nonabelian gauge fields at high temperature is
a spontaneous vacuum magnetization. It is closely related with asymptotic
freedom.
 In fact, asymptotic freedom at high temperature is
always accompanied by   the background stable, temperature
dependent and long range chromo(magnetic)
 fields  \cite{Skalozub1996}.
 The magnetization phenomenon  was investigated in detail in $SU(3)$ gluodynamics
\cite{Strelchenko2004} and  supersymmetric theories
\cite{Pollock2003}, \cite{Demchik2003}   by analytic methods and
in $SU(2)$ gluodynamics \cite{Demchik2008}, \cite{Antropov2014} by
the Monte-Carlo simulations on a lattice. In all these cases the
spontaneous creation of magnetic fields  has been detected. Within
application to the early Universe the spontaneous vacuum
magnetization in the electroweak sector of the standard model is
described in review paper \cite{Demchik2015}.

The case of experiments at the $LHC$ requires a  special
consideration. This is  because of much lower temperature $T_d $
as compared to the electroweak phase transition temperature $
T_{ew} \sim 100$ GeV. For temperatures $T_d < T < T_{ew}$ the
scalar field condensate, supplying particle masses, screens the
magnetic field $H$, which was generated at high temperatures by
the $W$ boson loops. At the same time, the color magnetic fields
$B_3, B_8$ remain unscreened \cite{Demchik2015}. Within this
scenario the question arises: whether or not there exists  a
mechanism generating magnetic field $H$ in between critical
temperatures $T_d$ and $T_{ew}$?

In a qualitative manner it was considered in our  paper
\cite{Skalozub2016}. Therein, in particular, we have demonstrated
that magnetic field $H$ can be generated due to the vacuum
polarization of quark fields by the constant color magnetic fields
$B_3$ and $B_8$, existed in the $QGP$ after the $DPT$. In the
effective potential of the external fields the mixing terms of the
type $\sim eH\times (g B_3)^3, \sim eH\times (g B_8)^3, etc$,
where $e, g$ are electromagnetic and strong interaction couplings,
present and act as the sources for $H$. The field $H$ is
temperature dependent and occupying a large plasma volume as  the
fields $B_3$ and $B_8$.

In the present paper, we investigate in detail the creation in
$QGP$ of the magnetic fields $B_3, B_8, H$   at temperatures close
to the $DPT$ and estimate the field strengths. The proper time
representation is used. The one-loop plus daisy diagrams effective
potential of external fields  $V(B_3,B_8,H,T)$ accounting for the
gluons and $u-, d-$ and $s-$quarks at finite temperature is
calculated. This field configuration is stable due to the daisy
diagram contributions which cancel the imaginary terms presenting
in the one-loop effective potential of charged gluons
$V^{(1)}(B_3,B_8,T)$. For estimation of the field strengths the
asymptotic high temperature expansion, derived by Mellin's
transformation technique, is applied. As corollary of these
investigations we observe that strong color magnetic fields
$B_3,B_8,$ of the order $\sim 10^{18}-10^{19}$ G and usual
magnetic field $H \sim 10^{16}-10^{17}$ G are generated for
temperatures $T \sim 160 - 220$ MeV. The spontaneous magnetization
disappears at $T \sim 110 - 120$ MeV. This temperatures is
considered as the deconfinement temperature in the presence of the
fields. It is essentially lower that the one estimated without
magnetic fields.

The paper is organized as follows. In  next section we adduce the
one-loop effective potential of quarks $V^{(1)}_q(B_3,B_8, H,T)$.
In sect. 3 we present the one-loop contributions of gluons
$V^{(1)}_{gl.}(B_3,B_8,T)$ calculated in the high temperature
approximation, which is sufficient for the problem under
consideration. In sect. 4 the contribution of daisy diagrams is
calculated in brief and the dimensionless variables used in
numeric calculations are introduced. Then in sect. 5 the values of
the field strengths $B_3(T), B_8(T), H(T) $ are estimated for a
number of temperatures. Discussion of the results and conclusion
are given in the last section. Appendix A describes the details of
calculations of the quark zero temperature effective potential.
Appendix B includes information about the high temperature
expansion of the one-loop effective potentials $V^{(1)}_q,
V^{(1)}_{gl}$.
\section{Quark contributions to  one-loop effective potential}
In what follows, we consider the situation when temperature of
$QGP$ is not much higher than $T_d$. In this case, according to
\cite{Strelchenko2004}, the color magnetic fields $B_3$ and $B_8$
are spontaneously created in the gluon sector of $QCD$ because
color symmetry is restored. On the contrary, for the temperature
interval $T_d < T < T_{ew}$ the electroweak symmetry is broken and
$SU(2)$ constituent of usual magnetic field is screened by the
scalar field condensate. At temperatures $T > T_{ew}$ the
spontaneous generation of this field takes place also
\cite{Demchik2015}. Having this picture in mind we calculate the
one-loop quark effective potential $V^{(1)}_q(B_3,B_8,H,T)$ at the
background of all three fields.

To be in correspondence with the notations  of
\cite{Strelchenko2004}, \cite{Skalozub2016}  we present the
$SU(3)_c$ gluon field in the form
\be \label{field} A_\mu^a = B_\mu^a + Q_\mu^a, \ee
where $B_\mu^a$ is background classical field and $Q_\mu^a$
presents quantum gluons. We choice the external field potential in
the form $B_\mu^a = \delta^{a3} B_{3\mu} + \delta^{a8} B_{8\mu}$,
where $B_{3\mu} = H_3 \delta_{\mu  2} x_1$ and $B_{8\mu} = H_8
\delta_{\mu  2} x_1$  describe constant chromomagnetic fields
directed along third axis in the Euclidean space and $a = 3$ and
$a = 8$ in the color $SU(3)_c $ space, respectively. The field
tensor has the components: $F^{ext ~a}_{\mu\nu}= \delta^{a3}
F_{3\mu\nu} + \delta^{a8} F_{8\mu\nu}$, $~ F_{c12} = - F_{c21} =
H_c,
 c = 3, 8 $. We direct usual magnetic
 field also along third axis and choice its  potential in the
 form: $A_{\mu}^{ext} = H \delta_{\mu 2} x_1$.

We first calculate the quark spectrum in the presence of all these
fields \cite{Skalozub2016}. The corresponding Dirac equation reads
\be \label{Deq} (i \gamma_\mu D_\mu + m_f ) \psi^a = 0, \ee
where $\psi^a$ is a quark wave function, $a$ is color index, $m_f$
is mass of $f$-flavor  quark.  The covariant derivative describes
the interactions with external magnetic fields $H$ and $H_3, H_8$:
\be \label{D} D_\mu = \partial_\mu + i q_f |e| A_\mu^{ext}  + i g ( T^3
B_\mu^3 + T^8 B_\mu^8 ) \ee
where $T^3 = \frac{\lambda^3}{2}, T^8 =\frac{\lambda^8}{2}$ are
the generators of $SU(3)$ group, $\la^{3,8}$ are Gell-Mann
matrixes. Due to the choice of the potentials  we can  present the
quark spectrum as the sum of contributions of the following
external field combinations:
\bea  \nn \mathcal H^1_f &=& q_f |e|H+g \left(\frac{H_3}{2}+\frac{H_8}{2\sqrt{3}}\right),\\
\label{fields} \mathcal H^2_f &=& q_f |e|H+g \left(\frac{H_8}{2\sqrt{3}}-\frac{H_3}{2}\right),\\ \nn
\mathcal H^3_f &=& q_f |e|H-g\frac{H_8}{\sqrt{3}}. \eea
Here, $q_f |e|$ is electric charge of   $f$-quark. Each flavor energy
spectrum is given by the  known expression (see, for instant,
\cite{Akhiezer69} )
\be \label{spectrum} \epsilon^2_{i,n,\rho,f} = m^2_f + p^2_z + (2
n + 1) \mathcal H^i_f - \rho   \mathcal H^i_f, \ee
where $p_z$ is momentum along the field direction,  $\rho = \pm
1.$

 Vacuum energy is defined as the sum of the modes having
negative energy.  At finite temperature, in the imaginary time
formalism for fermions, it is reduced to the summation over
discrete odd   imaginary energies $p_4 = \frac{(2l + 1) \pi}
{\beta}$, $\beta = 1/T$ is inverse temperature
\cite{kala84-32-525}, \cite{Satz2012}. The result yields
\cite{Skalozub1996}
\bea \label{VT}\nn  V^{(1)}_q(T, H_i) =  \frac{1}{8 \pi^2} \sum_{f =
1}^6 \sum^3_{i = 1}\sum^{\infty}_{l = - \infty}(- 1)^l\\
\times\int\limits^{\infty}_0 \frac{ d ~s}{s^3} \exp (- m^2_f s -
\frac{\beta^2 l^2}{4 s}) \bigl[ \mathcal H^i_f s \coth ( \mathcal H^i_f s) - 1
\bigr]. \eea
This is expression of interest.  The term with $l = 0$ is the
vacuum energy $V^{(1)}_q(H_i)$. Different type asymptotic
expansions of \Ref{VT} are given in \cite{Skalozub2016}.
\section{Gluon contributions to  one-loop effective potential}
In this section, to give a self-contained presentation, we
describe in brief the one-loop contributions of gluons to the
effective potential. A detailed calculations are carried out  in
\cite{Strelchenko2004}. The Lagrangian of $SU(3)_c$ gluodynamics
is well known
\be\label{L} L = - \frac{1}{4} F^a_{\mu\nu}F^a_{\mu\nu} + L_{gf} +
L_{gh}, \ee
where $F^a_{\mu\nu} = \pa_\mu A^a_\nu - \pa_\nu A^a_\mu - g
f^{abc} A^b_\mu A^c_\nu $ is the field strength tensor,
 $f^{abc}$ are the group structure constants, $a = 1,
2,...,8$. In actual calculations, we use the decomposition of the
gauge field potential as in \Ref{field}. The metric is chosen to
be Euclidian for introducing the imaginary time formalism.  The
gauge fixing term in \Ref{L} is
\be \label{gft} L_{gf} = - \frac{1}{2} (\pa_\mu Q^a_\mu + g
f^{abc} B^b_\mu Q^c_\nu)^2, \ee
and $L_{gh}$ represents the ghost field Lagrangian. The components
$Q^a_\mu$ with $a = 1,2,4,5,6,7$ correspond to the color charged
gluons. In calculations it is convenient to use the "charged
basis" of gluons
\bea \label{Basis1} \nn W^{\pm}_{1\mu} ~ = ~ \frac{1}{\sqrt{2}}(Q^1_\mu \pm
i Q^2_\mu), \\
 W^{\pm}_{2\mu} ~ = ~\frac{1}{\sqrt{2}}(Q^4_\mu \pm i
Q^5_\mu),\\ \nn
W^{\pm}_{3\mu} ~ = ~ \frac{1}{\sqrt{2}}(Q^6_\mu \pm i
Q^7_\mu). \eea
In terms of these fields the spectrum of the charged gluons looks
like the spectrum of the spin one massless charged particle with
gyromagnetic ratio $\gamma = 2$. That is, we have to set in
\Ref{spectrum} m = 0 and put  $\rho = \pm 2$ for gluons and $\rho
= 0$ for ghosts. As the values of the $\mathcal H^i$ we have to
put $H = 0$ and use the combinations of the neutral components
$H_3$ and $H_8$ entering \Ref{fields}.

The detailed calculation of the one-loop effective potential for
 $SU(2)$  was carried out in \cite{Skalozub1996}, \cite{Skalozub2000}. It corresponds to
 each $SU(2)$ subgroup of the $SU(3)_c$ group.  These subgroups are related with the components $W^{\pm}_{r \mu}$,
 $r = 1,2,3$ of the basis \Ref{Basis1} and the corresponding
 combinations of $H_3$ and $H_8$ (for
more details see \cite{Strelchenko2004}):
 \bea \nn  B_{r=1,\mu} &=& B^{3}_\mu, \\
               \label{Hr}  B_{r=2,\mu} &=& \sqrt{\frac{3}{2}} B^8_\mu +
                 \frac{1}{2} B^3_\mu, \\ \nn
B_{r=3,\mu} &=& \sqrt{\frac{3}{2}} B^8_\mu -
                 \frac{1}{2} B^3_\mu. \eea
Just these fields enter the covariant derivatives related with the
$SU(2)$ subgroups, $D^r_\mu = \pa_\mu + i g B^r_\mu.$

The one-loop gluon contribution to the effective potential,
presented in the form similar to \Ref{VT}, reads
\bea\nn \label{VTg} V^{(1)}_g(T, H_r) =  \frac{1}{8 \pi^2}
\sum^3_{r = 1}\sum^{\infty}_{l = - \infty} \eea
\bea\times\int\limits^{\infty}_0 \frac{ d ~s}{s^2} e^{- i \mu^2 s}
\exp(i l^2 \beta^2/4 s)\Bigr[\frac{g H_r \cos(2 g H_r s)}{\sin(g
H_r s)} - \frac{1}{s} \Bigl], \eea
where $\mu^2 - i \epsilon, \epsilon \to 0$ is a  parameter playing
the role of the normalization point in the field. It is useful for
analytic continuations from the weak fields $gH_r \leq \mu^2$ to
the fields $gH_r \geq \mu^2$ when an imaginary part of the
effective potential is calculated.  The term with $l = 0$ gives
the zero temperature (vacuum) part and other terms describe the
statistical part.

In what follows, we take into consideration the gluon
contributions in the high temperature limit $T >> \\ (g H_r)^{1/2}
\geq \mu$, which is sufficient for our problem. The calculation of
$V^{(1)}_g(T, H_r)$ for this case was carried out by Mellin's
transformation technique described in Appendix B for the quark
effective potential \Ref{VT}. For gluons it is presented in
\cite{Strelchenko2004}, Eq. (10) :
\bea \nn \label{VTgas} &&V^{ as}_{g}(T, H_3, H_8)= \frac{H_3^2}{2}
+ \frac{11}{32} \frac{g^2}{\pi^2} H_3^2 \log[\frac{T}{\mu}]  \eea
\bea \nn -(g H_3)^{3/2} \frac{T}{3 \pi}+ \frac{H_8^2}{2} +
\frac{11}{16} \frac{g^2}{\pi^2} H_8^2 \log[\frac{T}{\mu}] \eea
\bea \nn -( \la_+^{3/2} + |\la_-|^{3/2}) (\frac{3}{2})^{3/4}(g
H_8)^{3/2} \frac{T}{3 \pi} \eea \bea \nn  - i \Bigl[(g H_3)^{3/2}
+ ( \la_+^{3/2} + |\la_-|^{3/2}) (\frac{3}{2})^{3/4}(g
H_8)^{3/2}\Bigr] \frac{T}{2 \pi}\eea
\bea+ O(g^2H_{3, 8}^2), \eea
where tree-level terms of $H_3$ and $H_8$ were added, \\$ \la_{\pm}
= 1 \pm \frac{1}{\sqrt{6}} \frac{H_3}{H_8}$ and $\mu$ is
normalization point. This expression contains the imaginary part
related with the lower state of the gluon spectrum $\epsilon^2_{n
= 0,\rho = 2} = p^2_z - g H_r$. It was realized already, this term
is exactly canceled by the imaginary term coming from the daisy
diagrams for charged gluons \cite{Skalozub2000},
\cite{Strelchenko2004}. This is the main reason for using this
approximation in further analysis.
\section{Contribution of daisy diagrams}
 As it is well known \cite{Kapusta1989}, at
finite temperature along with a one-loop effective potential we
have to take into consideration the so-called daisy diagram
contributions which account for the long distance correlations.
Graphically, this is a series of one-loop gluon diagrams with
infinite number of insertions of the polarization tensors taken at
zero external momenta, $\Pi(T, g, H_3, H_8)$.
 They have the order $ \sim g^{3/2}$ in coupling
constant and therefore must be included in the effective potential
after the one-loop terms. The two-loop contributions have the
order $ \sim g^{2}$. This order terms were neglected in
\Ref{VTgas}.

The calculations of the daisy diagram contributions coming from
charged gluons are described in \cite{Skalozub2000},
\cite{Strelchenko2004}. In  notation \Ref{Hr} the part given by
the unstable modes is
\bea \label{Daisy1} && V^{daisy}_{unst.} =  \frac{ T}{2
\pi}\sum_{r = 1}^3 (g H_r)\bigl[\Pi^{r}(H_r,T) - g H_r
\bigr]^{1/2}
\\ \nn && + i \Bigl[(g H_3)^{3/2} + ( \la_+^{3/2} + |\la_-|^{3/2})
(\frac{3}{2})^{3/4}(g H_8)^{3/2}\Bigr] \frac{T}{2 \pi}. \eea
Here, $\Pi^{r}(H_r,T)$ denotes the one-loop polarization tensors
of color charged gluons averaged over the ground (unstable) state
of the tree-level spectrum taken at $p_3 = 0:|n = 0, \rho = 2> $.
As we see, the imaginary parts in \Ref{VTgas} is exactly
 canceled by the one in \Ref{Daisy1}. Thus, the effective
potential $V^{(1) as}_{g} + V^{daisy}_{unst.}$ is real if
$\bigl[\Pi^{r}(H_r,T) - g H_r \bigr] > 0$.

Detailed calculations of the charged gluon polarization tensor
have been carried out in \cite{Strelchenko2004} (see also review
\cite{Demchik2015}). The most important for us is the temperature
and field dependencies   of it: $\Pi^{r}(H,T) \sim g^2 \sqrt{g H}
T$. Hence, the remaining after the cancelations part in the
effective potential is of the order $\sim g^2 g^{1/4} $. It is
smaller than the accuracy preserved  in \Ref{VTgas}. We can
conclude that the role of the unstable mode daisies consists in
the stabilization of the effective potential in the chosen
approximation in coupling $g$. All other terms are negligibly
small and have to be dropped. As concerns the contributions of the
neutral gauge field daisies, they have the order $\sim g^{5/2}$
and have to be dropped also \cite{Strelchenko2004}. Thus, the
consistent effective potential for gluons including the one-loop
plus daisies is real.

To carry out numeric calculations we use the dimensionless
variables for the effective potential, temperature and fields. We
consider the proton mass $m_p = 938.27208$ MeV as a reference
parameter and introduce the dimensionless variables:
\bea \label{variables1} && V^0_{q, g} =\frac{V^{(0)}_{q,
g}}{m_p^4},~ V^T_{q, g}=\frac{V^{(T)}_{q, g}}{m_p^4}, ~ \mu_f=\frac{m_f}{m_p},\\
\nn && h_{f,a}=\frac{\mathcal H^a_f}{m_p^2},~ \beta_p=
m_p\beta;~\omega_f=\mu_f\beta_p.\eea
 We also take into consideration three sorts of
quarks with the masses and electric charges
\bea \nn\label{qmass} m_u = 336  MeV,~ q_u = \frac{2}{3}|e|,\\
~ m_d = 340 MeV, ~q_d = -\frac{1}{3}|e|,\\
\nn ~ m_s = 486 MeV,~ q_s =-\frac{1}{3}|e| \eea
and the coupling values $\alpha_s =1,~ \alpha_e =\frac{1}{137},~ g=\sqrt{4 \pi},$\\ $~|e|= \sqrt{\frac{4 \pi}{137}}$. The dimensionless
field strengths are: $x=\frac{|e|H}{m_p^2},~
x_3=\frac{gH_3}{m_p^2},~ x_8=\frac{gH_8}{m_p^2}$. The field
combinations  in \Ref{fields} and \Ref{Hr} should be expressed in
term of them. In next section we present the results of the
calculations fulfilled for  a number of temperatures. Details of
calculations are placed in the Appendices.
\section{Estimate of the field strengths}
The total effective potential used in our investigation consists
of the one-loop  quark contribution  \Ref{VT} including the zero
temperature term $V^0_q$ with l = 0 and the statistical part
$V^T_q$ with $l \not = 0$ and the gluon contributions \Ref{VTgas},
\Ref{Daisy1} presented above. The calculation of the zero
temperature quark potential is given in Appendix A. To be in
correspondence with the gluon sector approximation, we apply the
high temperature expansion for quark sector also. The calculations
are given in Appendix B. The final expressions read

\bea \nn \label{VT0qas} V_q^0=\frac{x^2}{2e^2}+\frac{x_3^2}{2g^2}+\frac{x_8^2}{2g^2}+\frac{1}{8\pi^2}\sum\limits_{f}\sum\limits_{a=1}^3
\Biggl[\frac{1}{3}h_{f,a}^2 \\
\nn -2h_{f,a}^2\left(2\Gamma_1\biggl(\frac{\mu_f^2}{2h_{f, a}}\biggl)+\frac{\mu_f^2}{2h_{f, a}}ln \frac{\mu_f^2}{2h_{f, a}}+2\zeta'(-1)\right)\\
+\left.\frac{1}{3}h_{f, a}^2ln\frac{\mu_f^2}{2h_{f, a}}+\frac{1}{2}\mu_f^4ln\frac{\mu_f^2}{2h_{f,a}}-\frac{1}{4}\mu_f^4\right] \eea

and
\bea \nn \label{VTqas} V^{T ,
as}_q=\frac{1}{4\pi^2}\sum\limits_{f}\sum\limits_{a=1}^3\left[\frac{2}{3}h_{f,a}^2 \right. \\
\nn \left.\times\left[\frac{1}{2}\left(\gamma+ln(\frac{\omega_f}{\pi})\right)
+\frac{7\zeta'(-2)}{4}\omega_f^{2}+\frac{31\zeta'(-4)}{64}\omega_f^{4}\right]\right.\\
-\left.\frac{h_{f,a}^4}{90\mu_f^4}\left[-1+\frac{31\zeta'(-4)}{8}
\omega_f^{4}\right]\right] . \eea
Here, $\Gamma_1(x)$ is generalized Gamma function, $\zeta(x)$ is
  $\zeta$-function, the notations are given in \Ref{variables1}.

These two expressions plus \Ref{VTgas} and \Ref{Daisy1} are used
in the estimation of the magnetic field strengths. For doing so we
numerically solve the stationary equations
\be \label{statEq} \frac{\partial V(H,H_3,H_8,T)}{\partial H} = 0,
~\frac{\partial V(H,H_3,H_8,T)}{\partial H_{3, 8}} = 0 \ee
at  a number of fixed temperatures and obtain the roots
$h_{min}^i(T)$. If for a particular set of $h_{min}^i(T)$  the
total effective potential is negative, we have to conclude that
these magnetic fields are spontaneously generated. The results of
the calculations are presented in Table 1.
\end{multicols}
\newpage

\begin{center}
\begin{table}[h]

\begin{tabular}{|c|c|c|c|c|c|c|c|}
\hline
T, MeV&         $x_8$&      $x_3$           &$x$ $ 10^{-3}$&        $V$&            $H_8 10^{19} G$&        $H_3 10^{18} G$&        $H 10^{17} G$\\
\hline
120&        $0.463423$  &$0.0762514$&     $-0.0512118$&         $-0.0023921$&        0.589129&                0.96935&                    -0.0762115\\
140&        $0.783271$  &$0.131652$&     $0.0404301$&         $-0.00625276$&        0.995738&                1.67363&                    0.0601666\\
160&        $0.900931$  &$0.15203$&     $0.248859$&         $-0.011033$&        1.14531&                1.93269&                    0.370343\\
180&        $0.998792$  &$0.168842$&    $0.476241$&         $-0.0167341$&       1.26972&                2.14641&                    0.708724\\
200&        $1.09212$   &$0.184812$&    $0.727744$&         $-0.0235134$&       1.38836&                2.34943&                    1.083\\
220&        $1.18195$   &$0.200116$&    $1.00549$&          $-0.0314384$&       1.50256&                2.54399&                    1.49633\\

\hline

\end{tabular}

\end{table}

Table 1.The  values of the field strengths spontaneously generated
at chosen plasma temperatures
\end{center}
\begin{multicols}{2}
 In Table 1, in first column we show the temperature. The  next tree columns
  give the values of the dimensionless field strengths, the next one shows the behavior of the dimensionless effective potential.
 The field strengths [Gauss] are shown in the last three columns.

We have detected the negative values of the effective potential
for the stationary field strengths.   It means that magnetic $H$
and chromomagnetic $H_3, H_8$  fields have to be generated
spontaneously after the $DPT$ in $QGP$. If the temperature is
lower than 110 - 120 MeV, the effective potential value is close
to zero.  Hence,  within the high temperature approximation
adopted, we expect the background fields disappear and confinement
is realized. We see from Table 1 that the strength of the magnetic
field is two orders of magnitude less than the strength of the
colored fields and equals $\sim 10^{16} $G  at the $LHC$
experiment temperatures.

\section*{Discussion and conclusions}
 The most interesting observation of the above investigation is two
fold. Firstly, with temperature lowering the magnetic field
strengths are decreased. Secondly,  simultaneously  the value of
the effective potential in the minimum, being negative, increases
and tends to zero. Beginning from the value $V_{min} \sim - 0.02$
at $T= 200$ MeV it equals to  - 0.002  at $T= 120$ MeV, that is it
increases in one order. Such type  behavior detects that the
magnetic fields act to decrease  the $DPT$ temperature $T_d$.
Magnetized $QGP$ must be created at essentially lower temperature
as compared to the zero field case.

The $T_d$  lowering  has also been  observed already in
\cite{Fedorov}, \cite{Cosmai2007}, \cite{Bali2012} for the $DPT$
in applied external magnetic  fields. In \cite{Cosmai2007}, in
particular, it was found in  lattice calculations that the
temperature $T_d $ can even be reduced to zero for sufficiently
strong color magnetic fields. In contrast, here  we determined
similar behavior  for  the magnetic fields spontaneously created
in $QGP$. Hence, the $DPT$ has to happen at essentially lower
temperatures $\sim 110 - 120$ MeV. For these temperatures the
minimum value of the total effective potential is very close to
zero. In the used approach this means the  magnetic field
screening and color comfinement. Really, as we noted already,
 asymptotic freedom at high temperature has  always be
accompanied by  temperature dependent background  magnetic fields
\cite{Skalozub1996}.  Screening of these fields  reflects the
destroying of the asymptotic freedom regime and color comfinement
at low temperature. The spontaneously generated temperature
dependent macroscopic magnetic fields  are intrinsic constituent
of $QGP$ and the signals of the $DPT$. In short, deconfinement  is
always accompanied  by macroscopic long range chromo(magnetic)
fields.

Above we have applied the  approximation for the effective
potential including the one-loop plus daisies, which is real in
the leading order $\sim O(g^{3/2})$ in coupling constant. Here, we
 mention   a number of the mechanisms for the magnetic field
stabilization at finite temperature. In detail this problem was
investigated  by either analytic methods of field theory or
simulations on a lattice. It is  discussed in \cite{Demchik2015}.
Qualitatively, two factors act to stabilize
 vacuum. First is a so-called $A_0$ condensate related
with the Polyakov loop \cite{Starinets1994} which appears after
the  $DPT$. It enters the gluon spectrum of the type
\Ref{spectrum} in the form $\cdot\cdot\cdot + (g A_0)^2$ and  acts
in favor of eliminating the instability. Second are the radiation
corrections forming the magnetic mass of charged gluons
$\Pi^{\perp}_{ch.}(H,T) \sim g^2 \sqrt{g H} T$ and  having a large
positive real part \cite{Skalozub2000}, which also acts to
stabilize vacuum. These mechanisms have been proven also  in
lattice simulations \cite{Demchik2008}, \cite{Antropov2014}. So,
we stress again that the used  approximation for the effective
potential is consistent and reliable.

Now, let us compare the obtained results with that of in
\cite{Skalozub2016} where to clarify the role of the quark loop
effects the color magnetic fields $H_3(T)$ and $H_8(T)$ were
estimated from the effective potential of the gluon fields, only.
The comparison shown that the field strength $H(T)$ is $\sim 15$
per cent less in this approximation. For example, at $T = 200$ MeV
$x(h) = 6.24 ~10^{-4}$ whereas from the present results we
obtained $x(h) = 7.28~ 10^{-4}$. But qualitatively this is close.

As it follows from the obtained results, in $QGP$ strong
chromo(magnetic) fields of the order $H_{3,8} \sim 10^{18}-
10^{19}$ G and $H \sim 10^{16}- 10^{17}$ G must be present. This
influences all the processes happening and may serve as  the
distinguishable signals of the $DPT$. Due to magnetization, in
particular, all the initial states of charged particles have to be
discrete ones. This could modify the cross sections of particular
processes and detected in experiments. Moreover, the fields, as
well as the $A_0$ condensate, generate new type processes with
$C$-parity violation, which also could be the signals of the
plasma formation. Detailed consideration of these problems will be
reported elsewhere.



\section*{Appendix A}
The term with $l = 0$ in (6) is the vacuum energy $V^{(1)}_{q.
vac}( H, H_3, H_8)$. The well known expression  for it reads
\cite{Akhiezer69}
$$8\pi^2V^{(1)}_{q. vac,i,f}( H, H_3, H_8)=$$
\be\label{Int}=\int\limits_0^\infty
\frac{ds}{s^3}e^{-m_fs}\left[\mathcal H^i_f s \coth(\mathcal H^i_f
s)-1-\frac{1}{3}(\mathcal H^i_f)^2s^2\right]. \ee
Calculation of this integral can be done by using   the $\Gamma-$
and $\zeta-$functions:

$$\int\limits_0^\infty  t^{x-1} e^{-\alpha t} dt=\alpha^{-x}\Gamma(x),\quad Rex>0,\quad Re\alpha>0;$$
\be\int\limits_0^\infty
t^{s-1}e^{-vt}(1-e^{-t})^{-1}dt=\Gamma(s)\zeta(s,v),\ee $ \quad
Res>0,\quad Rev>0$.

Let us set in  denominator of \Ref{Int} $s^3 \to s^{3-\varepsilon}
$ and consider the limit $\varepsilon\to0$. We present (19) as the
sum of four integrals and rewrite them using (20). In this way we
get
$$8\pi^2V^{(1)}_{q. vac,i,f}( H, H_3, H_8)=$$
$$-(2\mathcal H^i_f)^2\left(\frac{m_f^2}{2\mathcal H^i_f}\right)^{2-\varepsilon}\Gamma(\varepsilon-2)$$
$$-\frac{(\mathcal H^i_f)^2}{3}\left(\frac{m_f^2}{2\mathcal
H^i_f}\right)^{-\varepsilon}\Gamma(\varepsilon)$$
$$+2(\mathcal H^i_f)^2\zeta(\varepsilon-1;\frac{m_f^2}{2\mathcal H^i_f})\Gamma(\varepsilon-1)$$
$$+2(\mathcal H^i_f)^2\zeta(\varepsilon-1;\frac{m_f^2}{2\mathcal H^i_f}+1)\Gamma(\varepsilon-1).$$

Then we make an expansion  in  series over $\varepsilon$ and use
$\Gamma_1$-function for calculation of the derivative
$$\zeta'(-1;x)=\frac{d\zeta(\varepsilon-1,x)}{d\varepsilon}\biggl|_{\varepsilon=0}=\Gamma_1(x)+\zeta'(-1),$$

$\Gamma_1(x)=\int\limits_0^xln\Gamma(y)dy+\frac{1}{2}x(x-1)-\frac{1}{2}xln(2\pi).$

As a result, we obtain Eq.(16) for vacuum energy.

\section*{Appendix B}
The general method for calculation of the high temperature
asymptotic  used
 in Secs. 2-3 has been developed in \cite{Skalozub1996, Weldon1982}.

For the high temperature expansion of the one-loop quark effective
 potential  we perform the following steps. First, the expression in the brackets in (6) we expand in series over $s$ near the point $s=0$ and
 insert
  into the integral over s. First two terms of interest are
%
%
$$4\pi^2V_{i,f}=\sum\limits_{l=1}^\infty(-1)^l$$
$$\times\int\limits_0^\infty ds e^{-m_f^2s-\frac{\beta^2l^2}{4s}}\left(\frac{(\mathcal H^i_f)^2}{3s}-\frac{1}{45}(\mathcal H^i_f)^4s\right).$$

 Then we integrate over $s$ by means of the well-known formula for K-function
$$\int\limits_0^\infty ds s^{n-1}e^{-as-\frac{b}{s}}=2(\frac{a}{b})^\frac{n}{2}K_n(2\sqrt{ab}).$$

As a result, we can separate the sum over $l$ from the fields
$\mathcal H^i_f$
$$4\pi^2V_{i,f}=\frac{2}{3}(\mathcal H^i_f)^2\sum\limits_{l=1}^\infty(-1)^lK_0(m_f\beta l)$$
$$-\frac{(\mathcal H^i_f)^4}{90m_f^4} \sum\limits_{l=1}^\infty(-1)^l(m_f\beta l)^2K_2(m_f\beta l).$$

 For the next sums, we can calculate the asymptotic expressions by using Mellin's transformation  \cite{Skalozub1996, Weldon1982}
$$\sum\limits_{n=1}^\infty(-1)^n K_0(\omega n)=\frac{1}{2}\left(\gamma+ln(\frac{\omega}{\pi})\right)$$
$$+\sum_{n=1}^\infty
\frac{(2^{2n+1}-1)\omega^{2n}\zeta'(-2n)}{2^{2n}(n!)^2},$$

$$\sum\limits_{n=1}^\infty(-1)^n(\omega n)^2 K_2(\omega n)=-1$$
$$+\sum_{n=2}^\infty \left(\frac{\omega}{2}\right)^{2n}(2^{2n+1}-1)\frac{4}{n!(n-2)!}\zeta'(-2n).$$

As a result, we obtain the high temperature expansion of the
 effective
 potential (17).
The high temperature expansion of the gluon one-loop effective
 potential and daisies  has been done in
 \cite{Strelchenko2004}.

\end{multicols}

\end{document}